\documentclass{medphyspaper}
\usepackage{graphicx}
\usepackage{amsmath}
\usepackage{amsfonts}

\usepackage{csquotes}

\usepackage{amssymb}
\usepackage{siunitx}
\usepackage{cleveref}
\usepackage{booktabs}
\usepackage{multirow}
\usepackage{adjustbox}
\usepackage{comment}
\usepackage{float}

\newif\ifblind

\blindfalse

\newcommand{\ITWM}{\ifblind \textit{[Institution name hidden for review]} \else Fraunhofer ITWM \fi}

\ifblind
\else
\author[DKFZ,HIRO, HDPHYS]{Remo~Cristoforetti}
\ead{remo.cristoforetti@dkfz-heidelberg.de}

\author[ITWM]{Philipp~Süss}
\ead{}

\author[DKFZ,HIRO, HDPHYS]{Tobias~Becher}
\ead{}

\author[DKFZ,HIRO]{Niklas Wahl}
\ead{n.wahl@dkfz-heidelberg.de}

\address[DKFZ]{Department of Medical Physics in Radiation Oncology, German Cancer Research Center -- DKFZ, Im Neuenheimer Feld 280, 69120 Heidelberg, Germany}

\address[HIRO]{Heidelberg Institute for Radiation Oncology -- HIRO, Im Neuenheimer Feld 280, 69120 Heidelberg, Germany}

\address[HDPHYS]{Faculty of Physics and Astronomy, Heidelberg University, Heidelberg, Germany}

\address[ITWM]{Fraunhofer Institute for Industrial Mathematics (ITWM), Kaiserslautern, Germany}
\fi
\version{1}

\title{Trading robustness: a scenario-free approach to robust Multi-Criteria Optimization for Treatment Planning} 

\begin{document}
\maketitle
\begin{abstract}
\noindent\textbf{Background:}
Treatment planning in radiotherapy is inherently a multi-criteria optimization (MCO) problem, as it requires balancing competing clinical goals. Traditionally, the treatment's robustness is not formulated as a part of this decision making problem, but dealt with separately through margins or robust optimization. 

\noindent\textbf{Purpose:}
This work facilitates integration of robustness into multi-criteria optimization using a recently proposed efficient \enquote{scenario-free} (s-f) robust optimization approach: Utilizing variance reduction objectives, whose computation is independent of the number of chosen error scenarios,  robustness can become part of the multi-criteria decision making process at minimal computational overhead.

\noindent\textbf{Methods:}
The s-f approach relies on the fast evaluation of the expected dose distribution and mean variance during optimization independent of the scenario number. This is achieved by precomputation of expected dose influence and total variance influence matrices, which can then be used for repeated solving of subproblems in the two explored MCO approaches:  Lexicographic Ordering (LO) and full Pareto Front (PF) approximation. Different prioritization strategies within the LO approach are used to assess the impact of variance reduction on the optimization outcome. A 3-objective PF approximation, including a variance reduction objective, is generated to visualize and analyze trade-offs between the competing objectives.

The robust optimization is performed including \num{100} scenarios modeling setup and range errors, as well as organ motion, on 3D- and 4DCT lung cancer patient datasets. Robustness analysis is performed to assess and explore the efficacy of all optimization strategies.

\noindent\textbf{Results:}
The s-f approach enabled robust optimization in MCO with computational times comparable to nominal MCO. Both MCO strategies highlighted the interplay between dosimetric and variance reduction objectives. The LO approach showed how prioritization affects plan quality and robustness, while the PF analysis revealed a clear trade-off between robustness and organ-at-risk sparing.

\noindent\textbf{Conclusions:}
The proposed s-f robust optimization approach allowed the efficient application of robust MCO by significantly reducing the required computational time. The reported analysis highlighted the conflicting trade-off nature of plan robustness and dosimetric quality, demonstrating how robust MCO supports a more informed and flexible decision-making process in treatment planning.
\end{abstract}
\section{Introduction}
The primary objective of radiotherapy treatment planning is to achieve an optimal balance between clinical goals, including adequate coverage of the target volume and minimal exposure of the surrounding organs at risk (OARs). As clinicians must navigate trade-offs between these competing clinical goals, the treatment planning problem inherently represents a multi-criteria decision making problem. In treatment planning, this problem has been traditionally approached by representing the trade-offs as a weighted sum of objectives to be optimized with scalar optimization techniques. In practice, this often results in a tedious iterative tuning process to find clinically acceptable plans whose quality further depends on the planners' experience \cite{kierkels2015,bohsung2005}. To systematize the navigation of the planning trade-offs, more sophisticated multi-criteria optimization (MCO) approaches have been introduced into clinical workflows \cite{zarepisheh2022, hong2008, breedveld2019,teichert2019,breedveld2009,craft2012,thieke2007}. These methods offer advantages over traditional weighted-sum approaches by enabling consistent and automated navigation of a priori goals or by allowing a fast and interactive a posteriori navigation of available solutions \cite{breedveld2009,breedveld2019}. Common examples for a priori methods are hierarchical or lexicographic optimization\cite{zarepisheh2022,breedveld2012}, while the most common a posteriori method relies on Pareto surface approximation \cite{hong2008,craft2012}.

Beyond the challenges of decision-making, the treatment plan quality can be significantly affected by uncertainties such as anatomical variations, patient setup or range errors\cite{lomax2008,lomax2008a, liebl2014, kinoshita2025}. According to the specific radiation modality and treatment site, robust optimization techniques can effectively reduce the required margin size and the integral dose to the patient \cite{fredriksson2012a, unkelbach2018}.

However, traditional robust optimization approaches involve evaluating multiple error scenarios during both dose calculation and optimization which makes them computationally expensive. At the same time, standard MCO methods, typically require the sequential generation of multiple plans, exacerbating the computational complexity of the problem and scaling the optimization time and memory footprint to unfeasibility.

Conventional robust optimization algorithms rely on well defined mathematical metrics such as max value operators, for min-max optimization approaches or expected value operators for probabilistic planning \cite{unkelbach2018, fredriksson2012a}. Despite their rigorous formulation, predicting the impact of these operators on the final plan remains challenging, and their effectiveness closely relates to the accuracy of the uncertainty model employed. These operators influence the problem formulation in binary manner, i.e once applied, they are enforced fully on all solutions.

The choice of the robustness operator or margin dictates the nature of the optimized plan, which may become too optimistic or too conservative depending on the settings \cite{fredriksson2012a,fredriksson2014,zaghian2017}. Because of this rigidity, the influence of robust optimization on the final dose distribution can only be adjusted by adapting the model describing the uncertainty source and its parameters \cite{badiu2024, tarp2024, vandewater2016}, or, when possible, by adopting balancing strategies \cite{sevilla2025}. 

Two fundamental limitations thus hinder the integration of traditional robust optimization with Multi Criteria Optimization (MCO). The first is a practical limitation posed by the computational cost of a full scenario based approach. The second is a conceptual limitation posed by the rigid application of robustness operators, which prevents the direct and flexible control over how robust optimization influences the final solution.

Several approaches tackled the first practical limitation in both optimization and robustness analysis. Examples include Polynomial Chaos Expansion \cite{perko2016,hoogeman2015} and Analytical Probabilistic Models (APM) \cite{bangert2013b, wahl2020}, which utilize analytical methods to propagate uncertainty through dose calculation. These techniques efficiently estimate the first moments of the dose for each voxel, and eliminate the explicit calculation of all error scenarios during optimization. However, they often still entail significant computational costs associated with constructing the analytical model itself.

A solution to overcome the second limitation, and enable robust MCO, was proposed in literature \cite{chen2012b} where traditional robust operators are applied to selected objectives. A combination of robust and non robust objectives builds the Pareto front approximation allowing the control of plan robustness through trade-off between these objectives. This approach offers a viable solution to the conceptual limitations of robust optimization, but still bares the computational burden of explicit individual scenario evaluation during optimization.

As an alternative approach, the APM framework itself suggests incorporating variance reduction objectives alongside standard dosimetric ones in the optimization problem formulation. In this case robustness is quantified through variance as a well defined metric, and supports a convex optimizable objective function.

Variance can be estimated during optimization through the explict evaluation of all error scenarios, nonetheless baring again the same computational cost. Alternatively, the recently proposed \textit{scenario-free} approach to robust optimization enables the natural and efficient integration of mean variance reduction objectives \cite{cristoforetti2025}. This efficiency arises from precalculation of probabilistic quantities over the set of error scenarios, avoiding explicit evaluation of each scenario during optimization. 

Including variance reduction objectives in MCO using the scenario-free approach can potentially address both limitations. It enables explicit flexible control over dose variance at minimal computational cost, effectively introducing robustness as a criterion in the decision making process.

In this work we propose and investigate the application of the scenario-free approach to two MCO techniques, namely a Lexicographic Ordering (LO) and a Pareto front approximation and navigation approach. The LO strategy is a so-called two-phase $\epsilon$-constrained (\textit{2p$\epsilon$c}) approach \cite{breedveld2009,breedveld2012} while the Pareto front approximation is obtained through a sandwiching algorithm \cite{rennen2009b,craft2006, bokrantz2013a}. Several optimization configurations are explored to investigate the impact of variance reduction as a decision making criterion for the different strategies.

\section{Materials and Methods}
\subsection{Multi Criteria Optimization}
For the purpose of this work, the MCO problem can be stated as
\begin{equation}
\begin{aligned}
    \operatornamewithlimits{min}_{\boldsymbol{x}}\ \vec{F}(\boldsymbol{x}) &= (F_{1}(\boldsymbol{x}), \dots , F_{n}(\boldsymbol{x}))\\
    \mathrm{s.t.}\qquad \boldsymbol{x} &> 0\\
    \qquad c_k(\boldsymbol{x}) &\leq 0
\end{aligned}\ ,
\label{eq:opt-prob}
\end{equation}
where $\boldsymbol{x}$ is the set of optimization variables, usually representing the fluence weights and $\vec{F}(\boldsymbol{x})$ are the competing cost functions in vectorized form: Each $F_{i}(\boldsymbol{x})$ component represents an objective in the treatment plan and constitutes a dimension in the objective function space. The problem also includes an arbitrary number of $c_k(\boldsymbol{x})$ constraints that limit the feasibility space. Solutions for which no $F_{i}$ can improve without degrading at least one other $F_{j}$ are called Pareto efficient, or Pareto optimal.

Solution of problem (\ref{eq:opt-prob}) through standard optimization algorithms requires a scalarization of the vectorized cost function $\vec{F}(\boldsymbol{x})$. The standard approach consists in performing a linear combination of the objectives and minimize $F(\boldsymbol{x}) = \sum_{i} w_{i} F_{i}((\boldsymbol{x}))$ upon assignment of relative importance weights $w_{i}$. The weight vector $\vec{w}$ is assigned a priori based on the clinical relevance of the associated objective. The minimal value of the cost function obtained this way corresponds to a single point in the objective space, thus to a single optimal plan, and its position on the Pareto surface is eventually determined by the weight vector itself. In practice, a clinically relevant plan is achieved heuristically tuning the importance vector $\vec{w}$.

\subsubsection{A posteriori approach: Pareto Front navigation}
This \enquote{trial-and-error} process of tuning weights, waiting for the next solution, and repeating with an adapted set of weights $\vec{w}$, may practically be replaced by an automated strategy to generate plans with different sets of weights $\vec{w}$. The method of choosing $\vec{w}$ can be formalized, for example using sandwiching algorithms, to adequately cover the Pareto front. Combined with a navigation technique to interpolate between generated points, the clinical decision is then performed a posteriori. The planner can explore the tradeoffs between objectives and their impact on the plan metrics by intuitive graphical user interfaces.

The accuracy of the surface approximation so of the interpolated plans, is proportional to the number of calculated points. Given the computational cost of optimizing a single plan, the set of $\vec{w}$ vectors should be appropriately chosen to achieve sufficient accuracy with minimal number of optimizations.

\subsubsection{A priori approach: Lexicographic Ordering}
A second approach to the scalarization of problem (\ref{eq:opt-prob}) consists in applying a lexicographic ordering strategy. Contrary to the Pareto front approximation and navigation, the clinical decision is performed a-priori by means of a well defined hierarchical ordering of the objectives which are addressed sequentially. Each objective is optimized individually and constraining all the other objectives to their previously achieved value.
Formally, problem \ref{eq:opt-prob} is scalarized as:
\begin{equation}
\begin{aligned}
    \operatornamewithlimits{min}_{\boldsymbol{x}}\ &F_{i}(\boldsymbol{x})\\
    \mathrm{s.t.}\qquad \boldsymbol{x} &> 0\\
    \qquad c_k(\boldsymbol{x}) &\leq 0\\
    \qquad F_{j}(\boldsymbol{x}) &\leq \epsilon_{j}\\
\end{aligned}\ ,
\label{eq:LO-scalarization}
\end{equation}
for all $j < i$ and $\epsilon_{j} = F_{j}(\boldsymbol{x^{*}})$ is the previously achieved value, i.e. $\boldsymbol{x^{*}}$ is the solution achieved for $j = i$. This way objective $F_{j}$ does not deteriorate upon optimization of objective $F_{i}$.

In particular, the (\textit{2p$\epsilon$c}) approach applied in this work performs the optimization in two steps, or phases. The first phase optimizes each objective up to a prescribed clinical goal $b_{i}$ assessing its reachability. During this phase $\epsilon_{j}$ in \ref{eq:LO-scalarization} is defined as:
\begin{equation*}
\epsilon_{j} = 
\begin{cases}
    b_{j} & F_{j}(\boldsymbol{x^{*}})\delta < b_{j}\\
    F_{j}(\boldsymbol{x^{*}})\delta & F_{j}(\boldsymbol{x^{*}})\delta \geq b_{j}
\end{cases}
\end{equation*}
where $\delta$ is a slack variable introduced for numerical stability.
The second phase optimizes each objective to their best and tunes the optimality of the solution. In this phase, all other objectives are constrained, the index $j$ spans $\{1, \dots, n\}\backslash i$ and $\epsilon_{j} = F_{j}(\boldsymbol{x^{*}})\delta$.

Both techniques allow to explore the interaction between the optimization objectives $F_{i}(\boldsymbol{x})$ and assess their impact on the optimized plan.

\subsection{The scenario-free approximation to robust optimization}
Traditional robust optimization approaches may be expressed as applying a robustness operator to the objective function $\vec{F}(\boldsymbol{x})$. Such operators usually rely on a set of individual error scenarios that need to be explicitly evaluated in each iteration during optimization. This implies significant computational overhead in the application of such robust optimization techniques as the optimization time for each problem to be solved scales with the number of included scenarios. The complexity of the problem scales thus with the number of scenarios, which is exacerbated in the case of MCO approaches requiring optimization of multiple subproblems.

Furthermore, the application of a robustness operator fully enforces attention to the modeled uncertainty. The effect of robust optimization on the dose distribution strongly depends on the nature of the operator used and whether the resulting plan is overly optimistic or overly conservative directly derives from this choice. In an MCO context, the optimization of robust objectives modifies the shape of the Pareto surface, but not the dimensionality of the objective space itself. An optimization trajectory in this space is thus limited to only access robust solutions when robust optimization is applied, and only non-robust solutions when it is not, preventing the exploration of intermediate solutions.

As an alternative approach to robust optimization, the scenario-free approach \cite{cristoforetti2025} enables efficient evaluation of the expected dose distribution $\mathbb{E}[\boldsymbol{d}]$ and mean variance $\mathbb{V}_{m}\left[\boldsymbol{d}(\boldsymbol{x})\right]$ during optimization. These quantities are estimated from precomputed probabilistic quantities through:

\begin{align}
\mathbb{E}[\boldsymbol{d}] & = \mathbb{E}[\boldsymbol{D}]\boldsymbol{x}, \quad \quad \text{with: } \mathbb{E}[\boldsymbol{D}] = \sum_{s} p_{s} \boldsymbol{D}^{s} \\
\mathbb{V}_{m}\left[\boldsymbol{d}\right] & = \frac{1}{N}\boldsymbol{x}^{T}\boldsymbol{\Omega}\boldsymbol{x}, \quad \text{with: } \boldsymbol{\Omega} = \mathbb{E}[\boldsymbol{D}^{T}\boldsymbol{D}] - \mathbb{E}[\boldsymbol{D}]^{T}\mathbb{E}[\boldsymbol{D}]
\end{align}

where $\mathbb{E}[\boldsymbol{D}]$ and $\boldsymbol{\Omega}$ are the expected-dose- and a variance- influence matrix, respectively. Both matrices are generated during dose calculation from the full set of error scenarios. This way no individual scenario information is retained during optimization. $p_{s}$ is a probability weight associated to each scenario $s$. It is determined by the probability distribution modeling the uncertainty source, together with the adopted sampling method.

With the scenario-free approach, conventional dosimetric objectives are evaluated on the expected dose distribution. The optimization cost function is formulated as a composite of the dosimetric objective and the mean variance term:
\begin{equation}\label{eq:cost-function}
    \mathcal{F}(x) = f(\mathbb{E}\left[\boldsymbol{d}(\boldsymbol{x})\right]) + \mathbb{V}_{m}\left[\boldsymbol{d}(\boldsymbol{x})\right]
\end{equation}

This combination is motivated by the formal equivalence to the traditional expected-value robust optimization approach when the least-squares objective function is considered.

However, the variance reduction term can be treated as a separate independent variance reduction objective. This enables its integration into the clinical decision-making process, with its full potential most effectively explored through an MCO approach.

Additionally, accumulating multiple variance-influence matrices on selected sets of error scenarios allows to separate the variance contribution of different sources of uncertainty. In an MCO framework this allows tradeoffs not only of variance against dosimetric goals, but also among variance goals for different uncertainty models.

\subsection{MCO Robust Optimization Strategies}
Integration of the scenario-free robust optimization approach with MCO allows for several optimization strategies to be designed. With the a priori LO approach, the importance of variance reduction is controlled through a predefined priority rank. The a posteriori Pareto front approximation instead, provides full flexibility to evaluate the impact of variance reduction on the plan.

\subsubsection{LO strategies}\label{section:LOstrategies}
The initial exploration of trade-offs with the LO approach is performed including only setup and range uncertainties. Three prioritization strategies are defined to investigate the role of variance reduction in the optimization process. A fourth strategy is introduced to address target motion uncertainty through a 4D variance reduction objective.

For the first strategy, both dosimetric and variance reduction objectives are combined at every priority level. This enables direct comparison with a non-robust, margin-based approach, in which the PTV is generated by expanding the CTV with an isotropic margin and only the nominal planning CT is considered. Such comparison allows the validation of the proposed method and the estimation of relative optimization times of the scenario-free robust LO.

The second strategy prioritizes dosimetric goals over variance reduction objectives, ensuring that the dosimetric quality of the plan is not compromised by the introduction of variance reduction. In this configuration, robustness represents an additional feature of the plan that is achieved on top of, rather than instead of, dosimetric quality.

The third strategy places target variance reduction above all other objectives, explicitly favoring robustness for target coverage.

For the fourth strategy two variance reduction steps follow the dosimetric optimization. The two objectives address two different sources of uncertainty, namely target motion/anatomical changes and setup and range errors. The separation is achieved by accumulating the $\boldsymbol{\Omega}$ variance-influence matrices on separate sets of error scenarios, each encoding the desired uncertainty source.

\subsubsection{Pareto Front Approximation}
A robust Pareto front approximation is generated considering three combined objectives: two dosimetric objectives for OARs sparing and one variance reduction objective for the target. This allows for a three-dimensional representation of the Pareto front to be visualized for descriptive purposes. The Pareto surface is obtained using the total expected- and variance-influence matrices accounting for the combined setup, range, and breathing motion uncertainties. Multiple points are sampled on the Pareto surface using the sandwiching algorithm to create a convex-hull approximation.

\subsection{Treatment planning application}
\subsubsection{Objective functions and clinical goals}
\Cref{tab:objective_funtions_LO} presents a collection of objective functions considered in this work. The selection of such objectives for each ROI was guided by the clinical interpretability of the goal values applied for the LO approach, as well as the intention to demonstrate that the method’s effectiveness is not limited to a specific set of objectives.

The variance reduction objective was formulated as minimization of the variance term in \cref{eq:cost-function}. The equivalent uniform dose (EUD) objective depends on the $n$ parameter as defined in \cref{tab:objective_funtions_LO}. This parameter encodes the volume effect of the specific organ and for parallel OARs, values of $n \approx 1$ are usually applied\cite{matuszak2016, AAPM1662012}. An higher value of $n$ increases the penalty and the attention of the optimizer for higher doses. In this work a value of $n = 3.5$ was chosen. This choice allows to maintain a clear clinical interpretation of the goal value associated to this objective while inducing a stronger reduction of higher doses in the lung. This choice additionally ensures the convexity of the EUD function \cite{choi2002}.

\begin{table}[h]
    \caption{Objective function definitions, relative parameters and prescriptions. The objective functions are applied in different configurations according to the specific optimization strategy applied. The $\hat{\cdot}$ operator applied to $d^{i}$ represents the \textit{mean} operator.}
    \centering
    \begin{tabular}{l l l l}
        \toprule
        \textbf{Objective Type} & \textbf{Functional form} & \textbf{Parameters} & \textbf{Goal}\\
        \midrule
        Max Dose & $d_{max} + log(\sum_{i} e^{\frac{d_{i} - d_{max}}{\epsilon d_{max}}})$ & $\epsilon = 1e-3$ & \SI{60}{\gray}\\
        EUD & $\left [\sqrt[n]{\frac{1}{N} \sum_{i}d_{i}^{n}} - d_{ref}\right ]^{2}$ & $d_{ref} = \text{\SI{0}{\gray}, } n = 3.5$ & [$\text{\SI{40}{\gray}}]^{2}$\\
        Mean Dose & $| \hat{d_{i}} - d_{ref}|$ & $d_{ref} = \ \: \text{\SI{0}{\gray}}$  & \SI{15}{\gray}\\
        Squared Deviation& $\frac{1}{N}\sum_{i}(d_{i} - d_{ref})^{2}$ & $d_{ref} = \SI{60}{\gray}$ & [$\text{\SI{1}{\percent}} \cdot d_{ref}$ $\text{\SI{}{\gray}}]^{2}$\\
        Variance reduction & $\mathbb{V}_{m}[d_{i}] - v_{pres}$ & $v_{pres} = \ \text{\SI{0}{\gray}}$ & \SI{1}{\gray\squared}\\
        \multirow[c]{2}{*}{EUD + Sq. Overdosing} & $\sqrt[n]{\frac{1}{N} \sum_{i}d_{i}^{n}} - d^{EUD}_{ref} \ \ + $ & $d^{EUD}_{ref} = 0$& \\[2ex]
        & $\frac{1}{N}\sum_{i}(d_{i} - d^{o}_{ref})^{2}\Theta(d_{i} - d^{o}_{ref})$ & $d^{o}_{ref} = \: \text{\SI{20}{\gray}}$ & \\
        \bottomrule
    \end{tabular}
    \label{tab:objective_funtions_LO}
\end{table}

Different trade-offs between such objectives are explored using the aforementioned robust optimization strategies.

\Cref{tab:LOstrategies} in \cref{sec:Appendix} collects the priority lists designed for the strategies (Strategy 1-4). Variance reduction for the target structure is prioritized differently for each strategy described in \cref{section:LOstrategies}, balancing its relative importance against the dosimetric objectives for the OARs. For all strategies, an additional minimum dose constraint is applied to the target at \SI{95}{\percent} of the prescribed dose. The definition of target structure differs between the scenario-free and the nominal case. For the nominal case this is set to the PTV, while for the scenario-free approach it is always set to the CTV.

For the Pareto Front approximation, multiple objective functions for the same structure are combined and treated as single objectives (\cref{tab:objective_funtions_LO}). For the OARs, this includes a least-squares deviation on the EUD plus a squared overdose objective on dose, forming a single objective. A variance minimization term is applied on the CTV. Additional minimum and maximum dose constraints are applied to the CTV to ensure coverage and avoid hot spots. The minimum and maximum target doses are set to \SI{50}{\gray} and \SI{55}{Gy}.

\subsubsection{Patient datasets}\label{sec:Patient datasets}
In this work, a lung cancer patient dataset was selected from the cancer imaging archive \cite{data4DLung2016}. The dataset consists of a 4D-CT scan with \num{10} phases representing the breathing motion. This choice allows for the separate investigation of the different considered sources of uncertainty, particularly of setup and range uncertainties as well as anatomical motion.

For the investigation of setup and range errors, only the nominal planning CT, corresponding to the deep breath-hold position, was considered. Setup errors were modeled as rigid shifts of the isocenter in all three spatial directions, sampled from normal distributions with zero mean and a \SI{2.25}{mm} standard deviation. Range errors were modeled as relative and absolute deviations and sampled from normal distributions with zero mean and standard deviations of \SI{3.5}{\percent} and \SI{1}{mm}, respectively.

\num{100} combined error scenarios were generated to compute the probabilistic quantities. These quantities allow to estimate an expected dose distribution and a variance reduction objective referred to in the following as 3D-expected-dose and 3D-variance. This set of probabilistic quantities was used to investigate the LO optimization Strategies (1-3).

To address the uncertainty arising from breathing motion, a second $\boldsymbol{\Omega}$ matrix was computed including only the 4D-CT phases. The resulting variance reduction objective $\mathbb{V}_{m}^{4D}\left[\boldsymbol{d}\right]$ quantifies thus the uncertainty associated with breathing motion.

An expected dose influence matrix was also accumulated including the combined dataset of 3D error scenarios and 4D-CT phases. This matrix allows to estimate the expected dose of the combined dataset. The LO strategy 4 was applied considering this latter expected dose influence matrix for the application of the dosimetric objectives. The variance reduction objectives were instead applied through the $\mathbb{V}_{m}^{4D}\left[\boldsymbol{d}\right]$ and $\mathbb{V}_{m}^{3D}\left[\boldsymbol{d}\right]$ operators.

A third dataset was generated by combining setup and range uncertainties across all 4D-CT phases. In this case, \num{10} setup and range errors were sampled for each breathing phase, resulting in a total of \num{100} scenarios. The \num{10} setup and range error scenarios are the same for each of the \num{10} CT phases, assuming thus a full correlation between 3D scenarios across phases. Given the nature of the uncertainty sources this is closer to a realistic treatment setup where the setup and range uncertainty reflect the interfractional positioning variability and systematic over- and undershoot.
These scenarios were combined into a single expected dose influence and variance influence matrix, capturing the combined variation induced by all three sources.

All treatment plans were generated using the same irradiation setup, consisting of a three-field proton plan with gantry angles fixed at \SI{45}{\degree}, \SI{90}{\degree}, and \SI{135}{\degree}. A constant RBE value of \num{1.1} was assumed for all optimizations, and no rescaling of the dose distribution was applied. Additionally, the PTV margin for the non robust plan was set to \SI{4}{\milli\metre}.

\subsection{Robustness Analysis}
Robustness analysis was conducted for each optimized plan using the complete set of scenarios employed during the pre-calculation of the respective probabilistic quantities. In particular, the \num{100} 3D scenario pool was used to evaluate the robustness of LO strategies (1-4). For the forth LO strategy, the robustness analysis was additionally performed on the \num{10} 4D-CT phases dataset to investigate the impact of breathing motion alone, and on the combined total dataset of 3D and 4D scenarios, i.e the total \num{110} scenarios.
The robustness analysis for the Pareto surface was conducted for all sampled solutions on the set of correlated 3D and 4D scenarios.

Plan quality and robustness were evaluated through multiple metrics, including dose distributions, dose-volume histograms (DVHs), standard deviation distributions, and standard deviation volume histograms (SDVHs). An SDVH provides a relationship between the standard deviation of the delivered dose and volume. DVH points are also reported for a selection of plans. Unless otherwise indicated, the reported values refer to the median value of the DVH metric on the set of error scenarios and the relative inter quartile range (IQR).

These metrics provide a quantitative measure of uncertainty and a measurable impact of the optimization strategy. However, plan robustness is conventionally assessed by testing selected metrics against predetermined acceptance criteria, and a plan is considered \textit{robust} when the selected metric reaches the desired quality for a predefined percentage of scenarios.

When variance is minimized with the scenario-free approach, there is no rigorous mathematical way of deriving such confidence intervals from the scalar mean variance value estimated. Testing of the desired metrics against the scenario set needs to be performed for the selected solution, a posteriori and through robustness analysis.

Previous work \cite{cristoforetti2025} validated the proposed scenario-free approach against scenario-based probabilistic planning however, the proposed MCO integration shifts the focus toward decision making and robustness as a criterion. Conventional worst-case optimization serves thus as a more appropriate reference and is addressed by initial analysis.

The link between conventional robustness metrics and variance is investigated with the following strategy: a plan is initially optimized with a conventional composite-worst case optimization approach where the scalarized cost function with the maximum value among the error scenarios is minimized. A total of \num{28} setup and range error worst-case scenarios in addition to the nominal scenario are collected from the same probability distributions described in section \ref{sec:Patient datasets} according to Korevaar et al.\cite{korevaar2019}. The worst-case errors are sampled at $2\sigma$ covering $>$ \SI{95}{\percent} of the uncertainty space. The plan is optimized such that a CTV coverage of $V_{95\%} \geq 95\%$ is achieved for at least \SI{90}{\percent} of the worst-case scenarios. 

Three scenario-free plans are optimized using the same set of worst-case scenarios to compute the required probabilistic quantities. The objectives for the scenario-free plan are equivalent to the ones used for the worst-case optimization, with the only difference of being applied to the expected dose distribution as opposed to the individual scenario distributions. Additional variance reduction objectives are applied to the CTV, lung and heart. The relative importance of the variance reduction objectives is set and progressively relaxed for the three plans to: a) match the DVHs and SDVHs of the worst-case plan, thus obtaining an equivalent plan meeting the same acceptance criteria, b) obtain a less conservative result, closer to probabilistic optimization and still meeting the robustness criteria, c) further prioritize OAR sparing against target coverage but failing the acceptance criteria. The three s-f plans showcase how different priorities set for the variance reduction objectives can reproduce equivalent conventional results while providing the degree of freedom to explore solutions within the robustness acceptance range.

The robustness analysis for these plans and the following MCO strategies is performed on the corresponding, complete probabilistic scenario set used for optimization. The reported analysis is thus a probabilistic evaluation of the plans.
The chosen robustness criteria addresses target coverage as the number of scenarios for which at least \SI{95}{\percent} of the CTV volume receives at least \SI{95}{\percent} of the dose prescription ($V_{95\%} \geq 95\%$). The acceptance threshold for this analysis is arbitrarily set at \SI{90}{\percent} of the scenarios as well.

The passing criteria analysis is reported for the LO strategies alone, while the Pareto front analysis only focuses on the exploratory aspects of the proposed method with relative reduction of variance.

\subsection{Algorithm implementation}
Both the scenario-free approach and the LO approach were implemented in the open-source treatment planning toolkit matRad \cite{wieser2017}. matRad served as the development environment for individual scenario dose calculations, accumulation of probabilistic quantities, and the final robustness analysis performed on all obtained solutions.

In contrast, the sandwiching algorithm for Pareto front approximation and the corresponding navigation tool were developed by the \ITWM \cite{lammel2025,collicott2021} and incorporate the same probabilistic quantities computed in matRad.

\section{Results}
\subsection{Conventional robustness reference}\label{sec:wcReference}
A plan is initially optimized using the conventional worst-case optimization and the \num{29} worst-case scenarios set and compared against three s-f plans obtained relaxing the importance weight for the variance reduction objective. Robustness analysis is performed on the \num{100} randomly sampled set of 3D scenarios.

\begin{figure}[H]
    \centering
    \includegraphics[width=\textwidth, keepaspectratio]{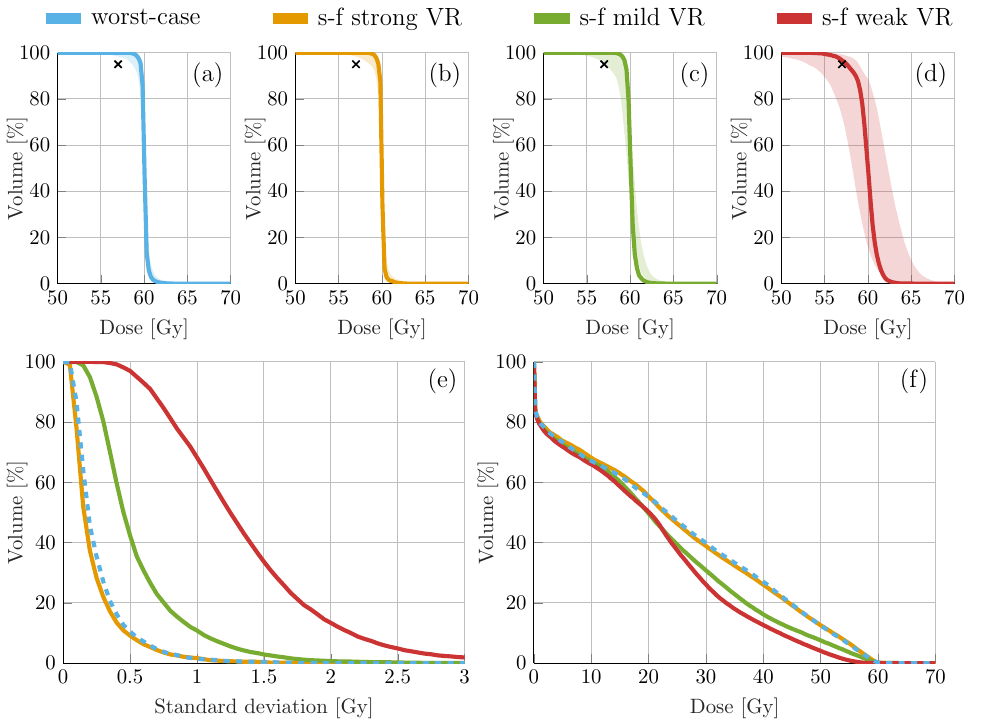}
       \caption{Probabilistic robustness analysis performed for the worst-case optimization and scenario-free approach in three different configurations. DVH and DVH band for the CTV structure and the worst-case (a, blue), equivalent scenario-free (b, orange) with strong variance reduction (VR), scenario-free with mild variance importance (c, green) and further reduced variance and target coverage importance (d, red). The mark reported on the DVH highlights the $V_{95\%} = 95\%$ point. SDVH for the CTV and all four plans (e). Lung DVH for all four plans (f). The blue line for the worst case and orange line for the scenario-free cases overlap significantly and are thus reported with different marks.}
    \label{fig:Figure_wcDVHAnalysis}
\end{figure}

\Cref{fig:Figure_wcDVHAnalysis} reports the robustness analysis performed on all plans. The worst-case solution and the s-f solution with tight variance reduction result in equivalent and overlapping DVHs, DVH bands and SDVHs for the CTV structure and the lung, practically resulting in equivalent plans fulfilling the same robustness acceptance criteria. The percentage of probabilistic scenarios meeting the robustness criteria ($V_{95\%} \geq 95\%$) is \SI{100}{\percent} for both plans.

The same analysis obtained for the mild- and weak-variance reduction scenario-free plans shows decreased target coverage and increased variability over the scenarios. Accordingly, the DVHs for the lung follow the opposite trend, showing higher sparing of the OAR with the increasing CTV variance.

The robustness criteria is met by \SI{100}{\percent} and \SI{63}{\percent} of the probabilistic scenarios for the mild- and weak-variance reduction plans, respectively. The mild-variance plan still satisfies the acceptance threshold set at \SI{90}{\percent} while the weak-variance reduction plan does not.

\subsection{LO approach}
All the reported results for the LO approach refer to the second phase of the \textit{2p$\epsilon$c} algorithm.
\subsubsection{Strategy 1: Algorithm validation}\label{Results:Validation}
\begin{figure}[H]
    \centering
    \includegraphics[width=\textwidth, keepaspectratio]{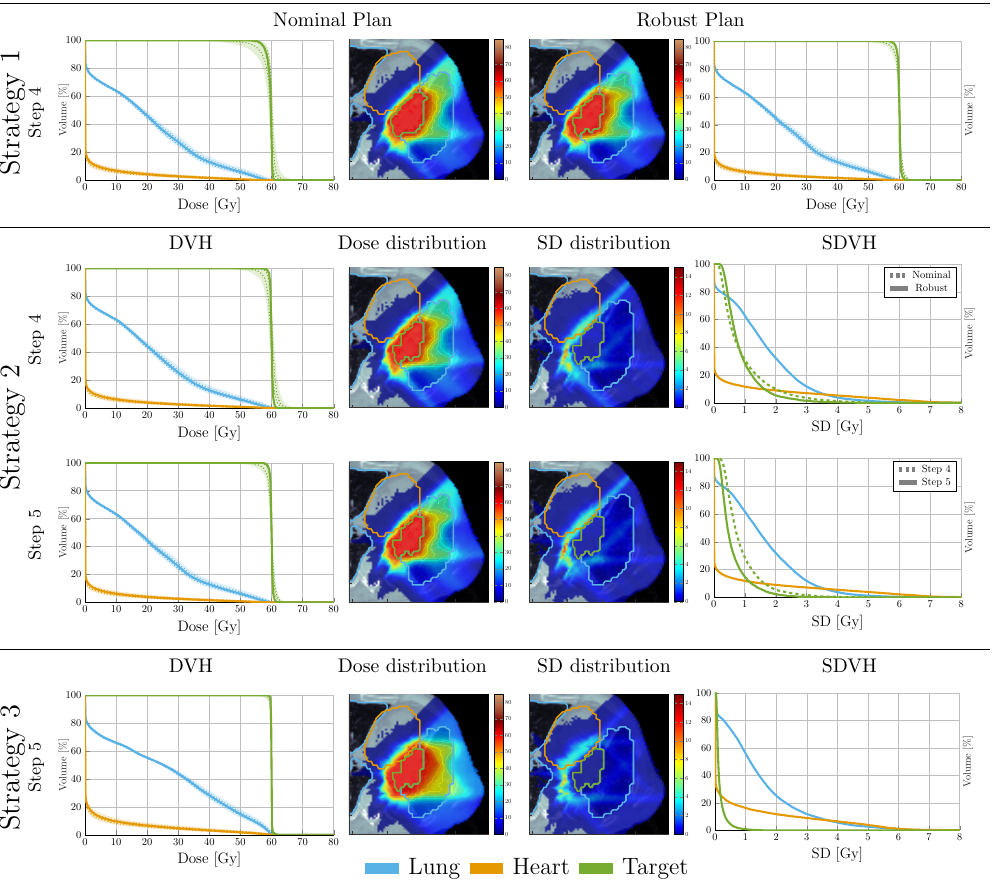}
       \caption{Comparison between optimized plans for \num{3} different LO optimization strategies. (Strategy 1, Step 4) Comparison between the nominal plan (left) and the scenario-free robust plan (right) for the last optimization step of the Strategy 1. (Strategy 2) Comparison between DVHs and expected dose distribution (left) and SDVHs and standard deviation maps (right) for the last dosimetric (Step 4) and variance reduction (Step 5) steps of Strategy 2. (Strategy 3) DVHs, expected dose distribution, SDVHs and standard deviation distribution of the last step of the optimization Strategy 3. For the DVH plots, the solid line corresponds to the DVH computed for the expected dose distribution, while the dotted lines correspond to the \num{25}-\num{75} percentiles. The colored DVH band spans the \num{5}-\num{95} percentiles over the scenarios distribution. For the SDVH of Strategy 2, solid lines for the target correspond to the SDVH for the corresponding step, strategy and plan, the dotted lines serve as a comparison reference for the nominal plan (Step 4) and the previous step (Step 5). All colorbar values are reported in \SI{}{\gray}.}
    \label{fig:Figure_1}
\end{figure}

The first set of results focuses on validating the scenario-free robust LO approach. \Cref{fig:Figure_1}, Strategy 1 presents a comparative analysis between the margin-based and the robust plans at the last step of the optimization sequence as reported in \cref{tab:LOstrategies}. The robust plan integrates both dosimetric and variance reduction objectives.

The DVHs include the target, left lung and heart structures. The reported solid line DVHs are computed from the expected dose distribution, while DVH bands span the \num{5}th to \num{95}th percentiles of the distribution of DVHs computed across all scenarios. A representative axial slice of the expected dose distribution is also reported for both algorithms.

The robust plan achieves a markedly improved robustness within the target, as demonstrated by the narrower DVH band relative to the nominal plan. In addition to improved target uniformity, the robust plan also results in better sparing of OARs. Notably, the integral dose to the lung is lower in the robust plan. The $V_{20}$ and $V_{30}$ values for the lung in the nominal plan are (\SI{45.9}{\percent}) and (\SI{27.3}{\percent}), respectively, whereas the corresponding values for the robust plan are (\SI{44.6}{\percent}) and (\SI{26.2}{\percent}), indicating a reduction in high- and mid-dose volumes within the lung. All the measured IQRs for these DVH points were below \SI{3.5}{\percent}.

The reported robustness analysis was performed with all \num{100} randomly sampled error scenarios used for the precalculation of the probabilistic quantities and is thus a probabilistic analysis. The percentage of scenarios meeting the robustness condition is of \SI{80}{\percent} for the margin based and \SI{92}{\percent} for the s-f robust plan. With the acceptance rate set at \SI{90}{\percent} the margin-based approach would thus not be accepted, as opposed to the s-f plan. 

Finally, the relative optimization times for the scenario-free approach range between 1.5 and 1.6 for all steps of the LO optimization sequence, meaning the scenario-free optimization is approximately 1.5 slower compared to the nominal optimization. Such overhead is introduced by the additional computation of the variance objective at each iteration of the optimizer. Total variance calculation is performed at each optimization step due to the total variance constraint applied in the second phase of the \textit{2p$\epsilon$c} approach. The rTPI refers to the optimization step alone and does not include the dose calculation step, which still needs to be performed for all scenarios. 
\subsubsection{Strategy 2: Prioritization of dosimetric objectives}\label{Results:DosePriority}

The second prioritization strategy focuses initially on dosimetric objectives alone. To highlight the impact of variance reduction, only the final dosimetric step (Step 4) and the first variance reduction step (Step 5) are reported for the robust plan in \cref{fig:Figure_1}. The robust plan at Step 4 in this analysis differs from the nominal plan presented for Strategy 1 (\cref{fig:Figure_1}) only in that dosimetric objectives are applied to the expected dose distribution rather than to the nominal one.
In Strategy 2,(\cref{fig:Figure_1}) the DVHs and expected dose distributions are shown on the left, while the standard deviation (SD) distribution and SDVHs are presented on the right.

At Step 4, the DVH plot still displays a relatively wide DVH band, particularly for the target, indicating that substantial uncertainty remains at this stage. This is further corroborated by the SDVH plot, where the SDVH for the nominal plan- corresponding to Strategy 1, Step 4 in \cref{fig:Figure_1}—is reported with a dashed line for reference. The SDVH for the robust plan is shown with the solid green line and differs only slightly from the nominal counterpart.

Variance minimization for the target structure is introduced in Step 5. The DVH band for the target becomes narrower, and the SDVH curve shifts to the left, indicating reduced standard deviation. For comparison, the SDVH for Step 4 of the same strategy is retained as a dashed line in the SDVH plot. No major changes are observed in the dose distribution between Steps 4 and 5. The SD distribution shows instead reduction of high standard deviation values and a more homogeneous pattern within the target.

The percentage of scenarios meeting the robustness criteria moved from \SI{80}{\percent} to \SI{89}{\percent} for Step \num{4} and step \num{5} respectively denoting the relation between variance reduction and improvement of conventional robustness metrics. The passing rate for the final plan would be close but not sufficient to meet the acceptance threshold set at \SI{90}{\percent}.


\subsubsection{Strategy 3: Prioritization of Variance reduction objectives}
The third prioritization strategy places variance reduction within the CTV volume as the primary objective. \Cref{fig:Figure_1}, Strategy 3 presents the DVH and SDVH analyses for the robust plan at the last steps of the optimization process. A highly robust dose distribution is achieved for the target as early as the first step of optimization and minimal variance is maintained throughout the remaining optimization steps.

The dose distribution reported for this strategy shows a lower target conformity and higher OARs doses compared to the results presented in section (\ref{Results:Validation}) and (\ref{Results:DosePriority}). At the same time the SD distributions and SDVH curves highlight how the standard deviation is highly reduced in the CTV compared to the previous approaches.

\SI{100}{\percent} of the scenarios meet the robustness criterion already at the first step of the optimization and this is maintained until the last optimization step.

\subsubsection{Strategy 4: 4D robustness}
\begin{figure}[H]
    \centering
    \includegraphics[width=\textwidth, keepaspectratio]{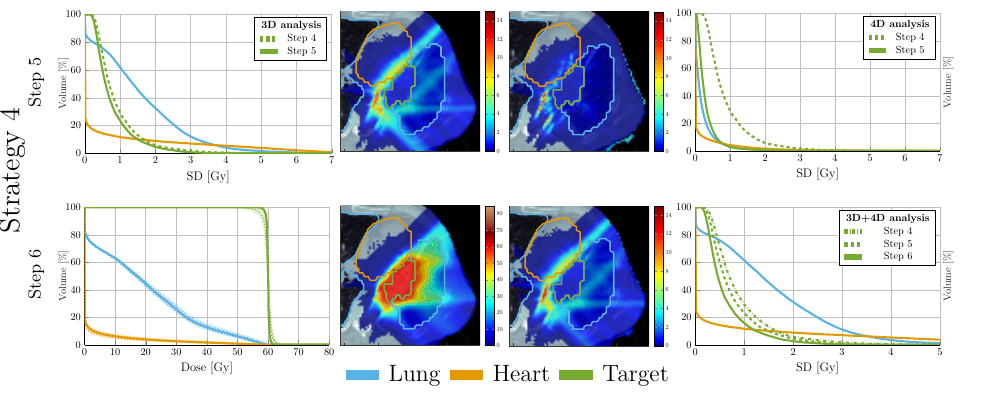}
    \caption{Robustness analyses performed for the Step 5 and Step 6 of the LO Strategy 4. Top left: SDVH analysis and SD distribution computed on the 3D-only set of error scenarios. The green dashed line corresponds to the target SDVH obtained at Step 4 for the same strategy and 3D error scenario dataset. Top right: SDVH and SD analysis for the same step but performed for the 4D scenario dataset  only including the nominal CT-phases. Bottom: DVH and expected-dose distribution (left) and SDVH and SD distribution (right) performed for Step 6 on the complete combined set of 3D and 4D error scenarios. All the reported values are in \SI{}{\gray}.}
    \label{fig:Figure_2}
\end{figure}

The fourth optimization strategy was designed to address setup and range (3D) uncertainties separately from breathing motion (4D) uncertainties. \Cref{fig:Figure_2} illustrates the robustness analysis associated with this strategy, focusing on the variance reduction achieved in Steps 5 and 6. In Step 5, 4D variance reduction is performed, and the robustness analysis was conducted on both the 3D-only and the 4D-only scenario datasets. For both robustness analyses, the figure also reports the SDVH for the CTV structure at the previous dosimetric-only step (Step 4) for direct comparison.

Both analyses present a reduction in variance for the target structure. The 4D analysis highlights a considerable reduction in target uncertainty introduced by the 4D-variance reduction objective. Concurrently, the 3D robustness analysis for the same Step also presents a contained reduction in target variance.

\Cref{fig:Figure_2} also presents the DVH and SDVH analysis for the final optimization step (Step 6), in which the 3D variance reduction is performed. At this stage, the robustness analysis was performed on the combined dataset of 3D and 4D scenarios, thereby accounting for the total uncertainty arising from both sources. For comparison, the SDVH plot also includes the curves obtained for the CTV, on the same dataset, in Steps 4 and 5, corresponding to dosimetric-only optimization and 4D-only variance reduction, respectively. Across these steps, the SDVH curve progressively shifts to the left, reflecting the reduction in uncertainty.

\subsection{Pareto Front approximation}
\begin{figure}[H]
    \centering
    \includegraphics[width=\textwidth, keepaspectratio]{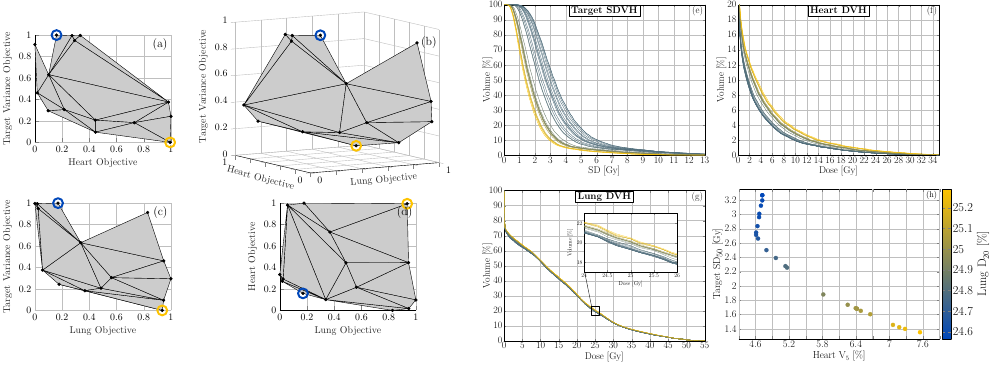}
    \caption{Representation of the Pareto front (a-d) for the considered case and correlation analysis for interpolated solutions on the surface (e-h). Black dots on the complete surface (b) and the three projections (a,c,d) correspond to points obtained through the sandwiching algorithm. The yellow and blue circles in (a-d) highlight Solution 1 and 2 respectively. SDVHs for the target (e), DVHs for the heart (f) and lung (g) for solutions obtained through interpolation of the Pareto surface. DVH ans SDVH lines with the same color correspond to the same solution. Scatter plot (h) relating $SD_{50}$ for the target and $V_{5}$ for the heart for the same set of sampled solutions. The set is color coded according to the $D_{20}$ metric for the lung of the corresponding solution.}
    \label{fig:Figure_Pareto_projections}
\end{figure}

A three-dimensional Pareto front approximation was constructed using the objective selection outlined in \cref{tab:objective_funtions_LO}. The resulting approximation, illustrated in \cref{fig:Figure_Pareto_projections}(b), represents the trade-offs among competing objectives within a normalized objective space [0,1]. The plans were sampled through the sandwiching algorithm, for a total of \num{15} solutions. Each solution is displayed as a black dot on the three-dimensional Pareto surface. To facilitate comparison among objectives, all bi-objective projections of the Pareto front are also shown (\cref{fig:Figure_Pareto_projections}(a,c,d)).
In addition to the surface representation, \cref{fig:Figure_Pareto_projections} reports a DVH and SDVH analysis for \num{25} solutions sampled on the Pareto surface. The solutions were sampled through convex interpolation of the surface along the target variance reduction axis. The solutions are color coded according to the objective function value for the target variance, from the lowest variance (yellow) to the highest (blue). A strong correlation between the objective values of the different solutions is observed, as the solutions with the lowest variance are the solutions with the worst DVHs for both the OARS.

\Cref{fig:Figure_Pareto_projections}(h) further highlights the correlation among solutions through a scatter plot for two selected DVH and SDVH metrics, namely $V_{5}$ for the heart and $SD_{50}$ for the target. $SD_{50}$ corresponds to the standard deviation value received by \SI{50}{\percent} of the target volume. To further highlight the correlation among the DVH metrics, the points are color coded according to the value of a third DVH metric, namely $D_{20}$ for the lung.

\begin{figure}[H]
    \centering
    \includegraphics[width=\textwidth, keepaspectratio]{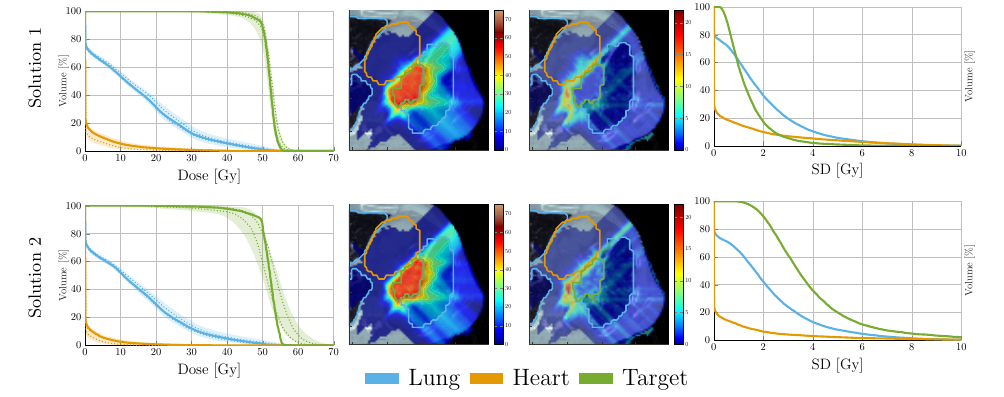}
    \caption{Comparison between DVHs and dose distribution (left), SD distribution and SDVHs (right) for Solution 1 (top) and Solution 2 (bottom). For the DVH plots, the solid line represents the DVH computed for the expected dose distribution, the dotted lines correspond to the \num{25}-\num{75} \SI{}{\percent} percentiles of the DVH distribution over the error scenarios, and the colored band corresponds to the \num{5}-\num{95} \SI{}{\percent} percentiles. All reported values are in \SI{}{\gray}.}
    \label{fig:Figure_Pareto_Solutions}
\end{figure}

\Cref{fig:Figure_Pareto_Solutions} reports the robustness analysis for two selected plans, chosen to highlight the extremes of the target variance dimension. The first plan, referred to as Solution 1, corresponds to the solution with the lowest variance reduction objective for the target. Its DVH band is tightly clustered around the expected dose distribution, indicating both high robustness and improved homogeneity within the target.

In contrast, Solution 2 represents the solution with the highest variance in the target. This plan exhibits a wider DVH band and reduced target homogeneity but achieves larger sparing of the OARs, with lower DVH curves for both the heart and lungs compared to Solution 1.

The SDVH curves provide additional insight into these differences. For the target, the SDVH in Solution 1 is significantly lower than in Solution 2, confirming its superior robustness. However, Solution 2 shows a lower SDVH for the heart, in accordance with the observed decrease in its DVH curve. The SD distribution for Solution 1 also reflects a more homogeneous and generally lower standard deviation within the target structure.

Both solutions are highlighted by yellow (Solution 1) and blue (Solution 2) circles on the Pareto surface in \cref{fig:Figure_Pareto_projections}(a-d).

From a multi-objective optimization perspective, Solution 1 is the second-worst plan in terms of heart sparing (Heart objective = \num{0.99}) and fourth-worst in terms of lung sparing (Lung objective = \num{0.94}).

When the Pareto front is projected onto the plane defined by the heart and lung objectives (\cref{fig:Figure_Pareto_projections}(d)), Solution 2 lies closest to the origin, i.e., it is the point with the lowest Euclidean norm. In other words, the solution with the best compromise between heart and lung objectives is the solution with the worst variance for the target.

These trade-offs can be explored interactively through interpolation of the Pareto surface, allowing for dynamic assessment of alternative solutions. This functionality is enabled by a graphical decision-support tool developed by the \ITWM. The tool includes real-time visual updates of objective projections and DVH comparisons.
An example of the use and functionality of this tool is available to the reader as additional material.

\section{Discussion}

The scenario-free approach incorporates variance reduction as a criterion in the decision making process. This enables explicit control of plan robustness through variance reduction, which is not bounded to any a-priori choice of robustness operator. Moreover, if the underlying uncertainty model does not accurately represent the real, undetermined, probability distribution of the error source, the scenario-free approach still provides a consistent solution to practically manage the impact of such uncertainty on the final dose distribution.

At the same time, the explicit incorporation of variance reduction terms into the cost function \cref{eq:cost-function} can strongly influence both the definition of optimization objectives and the optimization trajectory itself. Determining appropriate prescription values and the relative weighting of variance reduction objectives is not straightforward and poses a conceptual challenge.

Equivalency between scenario-free optimization and conventional robust optimization approaches may still be achieved by appropriate choice of the problem formulation. Previous work focused on the benchmarking of this approach against probabilistic optimization \cite{cristoforetti2025} and the analysis reported in \cref{sec:wcReference} provides a further comparison to worst-case optimization. This serves as an indicator that solutions equivalent to conventional robust optimization approaches are feasible for the s-f optimization as well. The range of equivalently robust solutions can be explored with the s-f approach accessing direct control on the mean variance of different structures the potentiality of which well suits a MCO approach.

There is no rigorous mathematical formulation allowing to derive acceptance criteria from non-local, mean variance values alone. Proper plan robustness can thus only be determined through a posteriori robustness analysis on a selected set of error scenarios.

However, it may be possible that robustness levels across patients remain comparable when similar goals on total variance are met. This would allow to determine reference or calibration points for future plans, facilitating more rigorous probabilistic planning with evidence-based upper bounds on variance ensuring acceptance in probabilistic evaluation after optimization.
The analysis in \cref{sec:wcReference} underlines this correlation between variance and conventional acceptance criteria, which is further supplemented by the remaining analyses of DVHs and plan metrics. 

Further, the ability to trade robustness a posteriori by navigating the Pareto front could actively help planners achieve the robustness goals. \citet{dejong2025} highlights the importance of performing probabilistic evaluation after planning to establish the compliance of the solution with the set robustness criteria. 
The s-f approach applied in a MCO framework instead would provide, if not directly the value of variance to meet the criteria, at least the objective direction in which to explore the space to improve the plan metrics.
\\

Thus, the contribution of this work is twofold. First, it demonstrates that robust MCO can be achieved with limited overhead compared to nominal MCO through the s-f approach, owing to its computational efficiency and flexibility. Second, by embedding variance reduction objectives within an MCO framework, this study enables systematic exploration of their effects on treatment planning, offering valuable insights into the behavior and impact of the s-f method.

To evaluate the performance and implications of the approach, two MCO strategies were applied and analyzed in detail.

\subsection{Prioritization strategies for the LO approach}

The first strategy combined dosimetric and variance reduction objectives at each optimization step. The resulting robust plan demonstrated a reduced sensitivity to uncertainty compared to the margin-based nominal plan and achieved improved sparing of organs at risk. Additionally, optimization times remained comparable to those of nominal optimization, supporting the feasibility of the s-f approach in clinical workflows. A direct comparison between the scenario-free robust and a traditional robust optimization algorithm in a multi-criteria optimization context is challenging due to the computational demand of the latter.

This first approach however, served as a validation of the proposed method. It reflects a more conventional clinical application of LO, where the order of objectives is inherited from the nominal optimization and the robust objectives are obtained by replacement of their nominal counterpart. In this configuration, prioritization is primarily driven by structure- and objective-based hierarchy rather than by a deliberate choice between dosimetric and robustness objectives.

Alternatively,  the s-f robust LO approach allows for the systematic exploration of alternative prioritization strategies that shift the balance between dosimetric and variance reduction goals. To investigate this, two contrasting strategies were considered. The first strategy (Strategy 2) prioritized dosimetric objectives and variance reduction is only performed at last, without compromising the achieved dosimetric goals. The second strategy (Strategy 3) reverses this priority by placing variance reduction at the top of the objective hierarchy.

For the first strategy (Strategy 2), a comparison between the robust and nominal plan at the end of the dosimetric phase effectively illustrates the difference between optimizing for expected versus nominal dose distributions. In the examined case, no substantial differences between the nominal and expected dose optimization were observed. It should be noted that all the presented results pertain to the second phase of the \textit{$2p\epsilon c$} approach, in which all objectives are constrained to their goal values or previously optimized outcomes. Consequently, the optimization of dosimetric objectives for the robust plan is anyway conducted under loose variance constraints, set here to an optimistic value of \SI{1}{\gray\squared}.

The subsequent variance reduction step effectively reduces the variance within the targeted structure thus proving the efficacy of the variance reduction objective in limiting the impact of uncertainty while still maintaining the achieved dosimetric plan quality. The robustness acceptance criteria set at $V_{95\%} \geq 95\%$ increased with the additional variance reduction step underlying the correlation between the value of such objective and the corresponding robustness metrics. For this particular case the acceptance threshold set at \SI{90}{\percent} was not met (only \SI{89}{\percent} was reached), meaning a more robust plan can be achieved increasing the variance reduction priority or relaxing the goal values for the OAR objectives with higher priority.

For the second strategy (Strategy 3), variance reduction alone applied at the first step would push the dose to the target toward zero. This is prevented by the hard constraint posed on the minimum target dose at \SI{95}{\percent} of the prescription.

This prioritization strategy yields a dose distribution for the CTV that fulfills the robustness criteria already at the first Step of the LO sequence and remains stable across all subsequent optimization steps. However, this robustness comes at the expense of target conformity and reduced sparing of nearby organs at risk, highlighting the trade-offs inherent in such prioritization.

Together, these two opposing strategies exemplify the flexibility afforded by the s-f LO framework. They demonstrate how variance reduction objectives can be treated on equal footing with dosimetric goals, allowing the prioritization configuration to be fully customized for each individual patient case. This opens the possibility for a broader spectrum of optimization strategies, tailored to clinical intent and case-specific trade-offs between robustness and plan quality.

The analysis in \cref{fig:Figure_2} illustrates how the scenario-free approach combined with lexicographic optimization allows the exploration of robustness against different sources of uncertainty.

This analysis demonstrates how the 4D variance reduction applied with Step 5 effectively reduces the uncertainty generated by breathing motion when it is assessed on the 4D scenario set. 4D variance minimization reduces the standard deviation for 3D scenarios as well, demonstrating how the two sources of uncertainty are not completely independent. Correlation among the uncertainty sources translates into correlation between the objectives and further demonstrates how variance reduction interacts as any other objective in the treatment plan.
For this patient case and the magnitude of the chosen setup and range errors, the uncertainty on the dose distribution introduced by breathing motion was lower compared to the uncertainty introduced by setup and range errors. This is reflected both by the standard deviation distributions and by the SDVH curves reported in \cref{fig:Figure_2}. 

For the analysis of Step 6, the uncertainty is estimated across all 3D and 4D scenarios and both variance reduction steps contribute to a partial reduction of the overall uncertainty.

Regardless of the specific prioritization strategy, each step in the LO sequence allows for a controlled relaxation of the previously optimized objectives. This is managed through the introduction of the slack variable $\delta$, \cite{breedveld2009}, which ensures numerical stability during the optimization process. No significant degradation in the quality of any previously optimized objective was observed across the analyzed cases, for any of the applied Strategies.

While the LO approach provides a structured and clinically interpretable prioritization of objectives, it effectively traces only a single path through the objective space and along the Pareto surface. In contrast, a full Pareto front approximation enables the exploration of the entire trade-off landscape between competing objectives, offering an extended view of the available planning solutions.

\subsection{Pareto front approximation}
The Pareto front representation and the reported analysis demonstrates how variance reduction, especially in the target, trades off against sparing of OARs.

Two extreme solutions along the target variance axis were highlighted. These exhibited significant differences in DVHs, DVH bands, SDVHs, and dose and standard deviation distributions. The most robust solution for the target (Solution 1) was among the least favorable for heart and lung sparing, while the best compromise plan for the OARs (Solution 2) presented the poorest target robustness. This inverse correlation reflects the optimizer’s behavior: robust target coverage requires expanding the high-dose region, thus reducing conformity and increasing dose to surrounding tissue; conversely, maximizing OAR sparing relies on a tighter dose distribution and steeper dose gradients, reducing thus robustness.

Correlation among objectives translates to correlation among DVH metrics as well. The analysis reported in \cref{fig:Figure_Pareto_projections}(e-h) for the interpolated solutions highlights how variability in the SDVH curves for the target translates to variability in the DVH curves for the two organs at risk. For this specific case, variability among DVH curves for the lung was lower than variability among DVH curves for the heart. The scatter plot reported in \cref{fig:Figure_Pareto_projections}(h) further enhances how metrics for target variance and organ sparing anti-correlate. When sampling the solutions along the variance reduction direction, intrinsic correlation among the OARs DVH metrics is also observed, as both of them improve when the variance objective worsen.

The analysis on the acceptance criteria is not reported for these solutions as the focus is posed on the opportunity offered by the method to explore diverse solutions. The correspondence with conventional robust optimization and the relation between variance reduction and passing criteria is already explored for the previous configurations and can easily be translated to this approach as well.

Combining this approach with the computational speed of the s-f framework allows efficient construction of the Pareto front approximation, effectively bringing robust Pareto front exploration within the reach of clinical use. The interactive decision-support tool further enhances this process by enabling real-time exploration of the Pareto surface, DVHs, and dose distributions, supporting informed clinical choices.

Although this analysis focused on three objectives only, realistic clinical scenarios involve higher-dimensional spaces. The extension of the approach to include additional dimensions poses no conceptual nor practical challenge. The sandwiching algorithm can handle the additional dimensions by sampling more solutions on the Pareto Front. Thanks to the s-f approach’s efficiency, the sampling of additional plans remains feasible within practical time constraints.

\subsection{Summary and Outlook}
The LO approach encodes precise clinical priorities directly into the problem formulation, offering a structured and guided optimization pathway. However, this inherently constrains the exploration of the objective space. In contrast, Pareto front approximation facilitates a broader exploration, allowing trade-offs among competing objectives to be fully visualized and evaluated. Depending on the clinical context, both strategies can support the decision-making process: some scenarios benefit from the clarity and structure of LO, while others may require the flexibility and comprehensiveness of Pareto-based exploration.

Hybrid strategies that combine the strengths of both approaches are also possible.  For instance, a lexicographic, clinically informed ordering could be applied to the dosimetric objectives, followed by a Pareto front approximation performed solely in the variance reduction space. This would allow robustness to be balanced across different structures, or different sources of uncertainty, without compromising initial dosimetric goals.
Additionally, the dual nature of the problem -- addressing both dose and variance -- suggests the potential for a bi-level optimization formulation. Recent work demonstrated the effectiveness of bi-level approaches for dose and LET optimization, particularly when supported by an efficient approximation of the domination cone \cite{schubert2025}. A similar formulation could be explored in the context of dose and variance optimization as well.

\section{Conclusions}
This work demonstrates the novel application of a scenario-free robust optimization approach to multi-criteria optimization. Successful integration into two MCO strategies --- a lexicographic ordering approach and a full Pareto front approximation -- was achieved. These approaches allowed to explore the impact of variance reduction objectives on the overall plan quality.

This approach effectively incorporates robustness as a criterion into the decision-making process, making it an integral, optimizable feature of the treatment plan. Notably, the observed optimization times remain comparable to those of a nominal MCO approach.

\ifblind

\else
\section{Aknowledgement}
NW acknowledges funding by the Deutsche Forschungsgemeinschaft (DFG, German Research Foundation), project no.~443188743. TB and NW acknowledge funding by the Cooperation Program in Cancer Research of the German Cancer Research Center (DKFZ) and Israel’s Ministry of Innovation, Science and Technology (MOST), project no.~Ca216.

\section{Conflict of interest}
The authors declare that they have no known competing financial interests or personal relationships that could have appeared to influence the work reported in this paper.
\fi

\section{Appendix}\label{sec:Appendix}
\begin{table}[h]
    \caption{Lexicographic ordering strategies applied to investigate the impact of variance reduction on the optimized plan. The strategies differ from each other in the prioritization grade of the variance reduction objective. The fourth and fifth column report the quantities on which the given objective functions are applied at the corresponding step. The minimum dose constraint was applied to all the strategies.}
    \centering
    \begin{tabular}{p{5pt} l l l c c}
        \toprule
        Constraints \\
        \midrule
        \multicolumn{2}{l}{\textbf{Structure}} & \textbf{Type} & \textbf{Parameters} \\
        \midrule
        \multicolumn{2}{l}{Target} & Min Dose & \SI{95}{\percent} $\cdot$ Prescription\\
        \\
        Objectives \\
        \midrule
        & \textbf{Priority} & \textbf{Objective Type} & \textbf{Structure} & \multicolumn{1}{c}{\textbf{Rob}} & \multicolumn{1}{c}{\textbf{ITV}}\\
        \midrule
        \multirow{4}*{\adjustbox{angle=90,valign=c}{Strategy 1}} & 1 & Max Dose & Heart & \multirow{4}*{$\mathbb{E}[\boldsymbol{d}]$} & \multirow{4}*{$\boldsymbol{d}$}\\
        &2 & EUD & Lung & &\\
        &3 & Mean Dose & Heart &  &\\
        &4 & Squared Deviation & Target & &\\
        \midrule
        \multirow{5}*{\adjustbox{angle=90,valign=c}{Strategy 2}} &1 & Max Dose & Heart & \multirow{4}*{$\mathbb{E}[\boldsymbol{d}]$} &\\
        &2 & EUD & Lung & &\\
        &3 & Mean Dose & Heart & &\\
        &4 & Squared Deviation & Target & &\\
        &5 & Variance & Target & $\mathbb{V}_{m}\left[\boldsymbol{d}\right]$&\\
        \midrule
        \multirow{5}*{\adjustbox{angle=90,valign=c}{Strategy 3}} & 1 & Variance & Target & $\mathbb{V}_{m}\left[\boldsymbol{d}\right]$&\\
        &2 & Max Dose & Heart & \multirow{4}*{$\mathbb{E}[\boldsymbol{d}]$} &\\
        &3 & EUD & Lung & &\\
        &4 & Mean Dose & Heart & &\\
        &5 & Squared Deviation & Target & &\\
        \midrule
        \multirow{6}*{\adjustbox{angle=90,valign=c}{Strategy 4}} &1 & Max Dose & Heart & \multirow{4}*{$\mathbb{E}[\boldsymbol{d}]$} &\\
        &2 & EUD & Lung & &\\
        &3 & Mean Dose & Heart & &\\
        &4 & Squared Deviation & Target & &\\
        &5 & Variance & Target & $\mathbb{V}^{4D}_{m}\left[\boldsymbol{d}\right]$&\\
        &6 & Variance & Target & $\mathbb{V}_{m}\left[\boldsymbol{d}\right]$&\\
        \bottomrule
    \end{tabular}
    \label{tab:LOstrategies}
\end{table}
\printbibliography
\end{document}